\documentclass[10pt]{iopart}

\usepackage{graphicx}
\usepackage{dcolumn}
\usepackage{bm}
\usepackage[usenames]{color}
\usepackage{hyperref}
\usepackage[normalem]{ulem}
\begin{document}

\title{Why it is so hard to detect Luttinger liquids in ARPES?}

\author{P.Chudzinski}

\address{Queen's University Belfast BT7 1NN, UK}
\ead{p.chudzinski@qub.ac.uk}
\vspace{10pt}
\begin{indented}
\item[]September 17th, 2018
\end{indented}

\begin{abstract}

The problem of photoemission from a quasi-1D material is studied. We identify two issues that play a key role in the detection of gapless Tomonaga-Luttinger liquid (TLL) phase. Firstly, we show how a disorder -- backward scattering as well as forward scattering component, is able to significantly obscure the TLL states, hence the \emph{initial state} of ARPES. Secondly, we investigate the photo-electron propagation towards a sample's surface. We focus on the scattering path operator contribution to the \emph{final state} of ARPES. We show that, in the particular conditions set by the 1D states, one can derive exact analytic solution for this intermediate stage of ARPES. The solution shows that for particular energies of incoming photons the intensity of photo-current may be substantially reduced. Finally, we put together the two aspects (the disorder and the scattering path operator) to show the full, disruptive force of any inhomogeneities on the ARPES amplitude.      

\end{abstract}

\section{Introduction}

Over the last three decades we have witnessed a significant experimental effort to find an unambiguous evidence of exotic non-Fermi liquid 1D states, in particular gapless Tomonaga Luttinger liquid (TLL). Angle Resolved Photo-Emission Spectroscopy (ARPES), a probe that can directly access the spectral function and hence reveal the spin and charge collective modes, has been frequently a method of choice. Various materials were chosen starting (more than 20 years ago) from organic Beechgard salts\cite{Zwick97, Vescoli2000, Claessen02} through  cupric-like oxides\cite{Denlinger99, Mizokawa02, Kim2006} until most recent studies of columnar 1D materials\cite{Watson17}, artificially created structures\cite{Schafer08} and domain walls\cite{Ma2017, Ohtsubo15} where we are able to cite only a tiny fraction of numerous papers on this subject. However this research field has proven to be extremely unrewarding, the TLL modes seemed to be unnaturally elusive: the signals were so weak that indirect methods of detection were proposed\cite{Gweon03}, the source of the putative power-law pseudo-gap has been frequently ambiguous as explained in detail in several recent review articles\cite{Grioni09, Dudy13, Dudy17}. One can then ask a question: does the nature of 1D states changes the conditions of ARPES experiment and makes them hard to detect? In this work we show that the answer to this question is affirmative.

Qualitatively, one can put forward a following argument. For a 3D state we need to account for each contributing atom as a site of possible emission/absorption/scattering and integrate over entire volume exposed to the light spot. The material is dense and moreover the 3D state can coherently re-adjust to accommodate the presence of a photo-electron, hence a homogeneous final state is favoured. In 1D the situation is different. Each 1D state acts as a separate source of photo-electrons. The initial state has no coherence along the directions perpendicular to 1D axis, we only observe a superposition of outgoing electronic waves (the respective final states). The key results of this work are based on this fundamental difference.

The aim of the current work is to show how the deep-rooted nature of 1D states can cause their ARPES intensities to be suppressed in comparison with a standard 3D system. The suppression effects that we capture here are different from standard, microscopic dipole matrix elements symmetry rules, as here we focus on mesoscopic phenomena. %In particular we will focus on two issues: disorder (in Sec.) and vertex corrections to photo-electron emission amplitude (in Sec.) The second effect will be also an effort to move from three-step towards a one-step description of ARPES in TLL, namely we devise a two-step approach to model the process. %in a one-step model and it reveals substantial dependence on incoming photon energy. Any experiment pertaining to access TLL must take these issues into careful examination. 
We will attempt to get an insight into the ARPES amplitude going beyond the commonly used three step model. A general idea behind its extension, towards the one step model, is to derive the full wavefunction of photo-electron inside the sample (the \emph{final} state) and compute its overlap with \emph{initial} states (i.e. eigenstates of N-1 fermions in a material under consideration). The task may be now naturally divided into two sub-problems: how the disorder affects the spectral function of TLL (Sec.\ref{sec:disord}) and how disorder affects the \emph{final state} (Sec.\ref{sec:rand-Lomel})  that will have been constructed in Sec.\ref{sec:Lommel}. In addition, in Sec.\ref{sec:Lommel} we show that, in a peculiar 1D case, the signal of TLL spectral function can be substantially reduced or enhanced depending on photon energy. Similar phenomena, related to the \emph{final} state interference effects, has been theoretically proposed\cite{Shirley-theo-interfer} and then experimentally explored\cite{Mucha-experim-bilayer, Isabellas-dark-corridor} (incl. photon energy dependence\cite{Ayria-PES-E-depend}) in quasi-2D graphene multi-layer systems, where the simplest version of a wave superposition takes place. In this respect our work is an extension that captures more complex diffraction patterns. 

\section{Statement of the problem}\label{sec:model}

\subsection{Initial state}\label{ssec:mod-initial}

We consider a material with negligible hopping in all but one direction, such that it can be described as a sum of 1D systems. Inside each system the \emph{initial state} of ARPES is described by the Tomonaga-Luttinger liquid hamiltonian:
\begin{equation}\label{eq:ham-TLL-def}
    H_{A}^{1D}[\nu]= \sum_{\nu} \int \frac{dx}{2\pi}
    \left[(v_{\nu}K_{\nu})(\pi \Pi_{\nu})^{2}+\left(\frac{v_{\nu}}{K_{\nu}}\right)(\partial_{x} \phi_{\nu})^{2}\right]+H_{cos}
\end{equation}

where $v_{\nu},K_{\nu}$ are velocity and TLL parameter
($\sim$compressibility) of a given bosonic mode $\nu$, these
depend on electron-electron interactions $V(r)$ with small momentum
exchange. This means they incorporate all terms
$V(q\rightarrow 0)$, the so called forward scattering terms $g_2$ and $g_4$. The terms
$V(q\rightarrow 2k_F)$, the so called back-scattering $g_1$ terms, enter as
additional cosine terms $H_{cos}$ and determine instabilities of the system. For concreteness, we consider here TLL with long range interactions. In this regime $g_4 \geq g_2 \gg g_1$ which means that the system may stay gapless down to very low energies, while at the same time the holon velocity can be much larger than Fermi velocity and $K_{\rho+}\ll 1$. $K_{\rho+}=1/[1+4V(q=0)],v_{\rho+}=1/K_{\rho+}$. Here we use the fact that the Coulomb like interactions involve only the symmetric charge mode $\phi_{\rho+}$. % and keep the Galilean invariance intact. %The TLL parameters for all other modes can be affected only by small $V_{ex}$. As long as the probability of spin/band exchange are approximately equal then $v_{rho-}=v_{\sigma+}=v_{\sigma-}$ and the three dispersions of corresponding bosonic branches are degenerate. 
The theory is written in terms of collective modes, the density fields
$\phi_{\nu}(x)$ and canonically conjugate fields $\theta_{\nu}(x)$, with
$\Pi_{\nu}(x)=\partial_{x}\theta_{\nu}(x)$, where $\nu$ is an
index of spin and charge modes. These fields are directly related to respective fermionic densities $\partial_{x}\phi_{\nu}(x)=-\pi \rho_{\nu}(x)$ which allows for immediate physical interpretation as a density fluctuation.  

The definition of density in reciprocal space $\rho_q= \sum_k c_{k+q}^{\dag}c_k$ reveals an intrinsic many body character of the TLL eigenstates which (if we move the sum on the l.h.s.) define our initial state $\Psi_i=\sum\Psi_{TLL}$. The \emph{initial} state of ARPES, the state of material after emission event, for 1D system is much more complicated to construct than in a 3D material because the entire spectral weight is shifted to the collective modes. Hence one needs to account interactions before evaluating the ARPES amplitudes. In physical realizations in quasi 1D materials, the 1D system is rarely defined one sole orbital of one sole type of atoms, it is usually an emergent state from a multi-site unit cell. Each the 1D collective mode is in essence an integral (over all available momentum) of DFT densities and one has to keep in mind that as interactions can modify occupations of various orbitals and inter-site hopping integrals, and exchange-correlation interactions are momentum dependent\footnote{and furthermore the un-occupied orbitals can be ad-mixed by interactions, see for example CI extensions of single determinant HF methods}, then it is effectively impossible to write down the expression in terms of single-electron orbitals. To by-pass this problem we take a hydrodynamic description of the TLL liquid with an averaged density with a cylindrical symmetry. Naturally, each 1D system does have internal orbital structure and it will determine the emission probability, however here we solely focus on the process of electron's propagation towards the surface assuming that standard dipole matrix elements has already been accounted for. One could state that we construct here a \emph{two-step} photo-emission model, where we separate emission of photo-electron inside a given 1D system and its later propagation (inside as well as outside the sample). 

\subsection{Final state}\label{ssec:mod-final} 

The final state is defined in a Fock space as a tensor product of a photo-electron $\psi_f$ and TLL with one electron missing $\Psi_f=\psi_f \otimes \sum\Psi_{TLL}^{N-1}$. Intuitively, one uses a sudden approximation\cite{Feibelman-ARPESbasic-theo}, where a photo-electron is immediately removed from the sample hence the two components are independent (mathematically the tensor product simplifies to an arithmetic product), however that would imply an infinite penetration depth of photo-electrons. We assume that the photo-electron does spend a finite amount of time inside the sample. Furthermore, it is customarily assumed that the 'leftover state' (TLL with one electron missing) is equivalent to an initial state $\Psi_{TLL}^{N-1}\equiv\Psi_{TLL}$. It is this approximation that allows one to relate ARPES signal directly to the spectral function of the initial state\cite{Feibelman-ARPESbasic-theo}. To be precise, one can define a weaker assumption: for any phonon flux that produces a noticeable photo-current, the \emph{initial state} is also not a pure, equilibrated TLL, but rather some steady state that is stabilized in the presence of a non-hermitian term in Eq.\ref{eq:ham-fin-nonherm}.  We keep this approximation, based on the fact that all gapless 1D states must belong to TLL universality class. However we note that for an extremely disordered system, where nanoscopic charging effects play a role, this is assumption does not hold.

The hamiltonian $H_f$, whose eigenstate is the photo-electron part of the \emph{final} state $\psi_f$, is non-hermitian:
\begin{equation}\label{eq:ham-fin-nonherm}
	H_f|\psi_f\rangle=(H_0+V_{ion}(r)+\imath\tilde{\Gamma})(r)|\psi_f\rangle = E_f |\psi_f\rangle~~,~~ \exists E_f \in \Re
\end{equation}
where $H_0$ is a free electron hamiltonian (with a discontinuity of chemical potential on the surface), $V_{ion}$ is a potential from a surrounding crystal lattice and the imaginary part $\tilde{\Gamma}$ of the potential is introduced by various inelastic scattering mechanisms. These incorporate all collisions that may be experienced by a photo-electron propagating towards the surface. In the following we will simplify the problem by taking a constant $V_{ion}(r)$ for $r>R_0$ where $R_0$ is a radius of 1D chains, while for $r<R_0$ is a trapping potential (quadratic in the simplest case). Similarly for the imaginary part $\tilde{\Gamma}(r)$ in the simplest approach one can distinguish losses taking place during collision event (for $r<R_0$) and when photo-electron moves through the environment ($r>R_0$). For the motion through the environment one usually takes an effective medium and forward scattering approximation where the source of losses are primarily due to interactions with electron hole pairs (dipoles and currents), so this is a term of the form $\tilde{\Gamma}_{out}\approx\vec{p}\vec{A_{ind}}$ with $A_{ind} \sim \psi(r_j)/\epsilon(r_i - r_j)$ (here $\psi(r_j)$ is the wave-function of the screening electrons and $\epsilon(r)$ is a dielectric function of the material). For $\tilde{\Gamma}_{in}$, which is much larger as it incorporates all intrinsic losses during the complicated creation of photo-hole event, we take it to be a constant parameter.  

The task is to solve the Schrodinger equation with hamiltonian $H_f$. The \emph{final} state $\psi_f$ is usually assumed in the following form:
\begin{equation}\label{eq:final-def}
\psi_f(r)=\sum_{r'} \sum_i \hat{\tau}(r,r') \phi_l(r'-r_i)
\end{equation}
where $\phi_l(r-r_i)$ is some complete basis set of wavefunctions defined on the $r_i$-th site. %Usually it is taken as atomic wave-functions with an underlying assumption that the \emph{initial} state is close to a single electron state(s). If this assumption holds then this simplifies calculation of local dipole matrix elements (see below). Here, for simplicity,
Here we will take a Hermite polynomials basis set.%, which is justified also by the fact that local dipole matrix elements are beyond the scope of our current work. 
The $\hat{\tau}(r,r')$ is called a scattering path operator and describes the motion of the high energy photo-electron towards the surface\cite{Osterwalder-multiple-scattering}. In the simplest mean field approximation of effective medium we assume that the distortion of all other electrons $\psi(r_j)$ is proportional to photo-electron density $\psi(r_i)$ hence the $\tilde{\Gamma}_{out}$ term introduces an exponential decay factor to $\psi_f(\vec{r})$ (this attenuation is well known as an electron escape depth, the fact that it factorizes out comes from the fact that it depends on variable $r_j$). The fact that $\tilde{\Gamma}_{out}\sim Im[1/\epsilon]$, that is electron energy loss function, is in full agreement with standard theory of escape depth\cite{Norman-escape-depth}, in the particular case of TLL (which is an under-screened metal) we predict smaller looses (limited phase space of electron-hole excitation) hence larger escape depth.  

On its evolution the $\psi_f$ may undergo typical inelastic scattering/absorption on various states present in the solid (e.g. phonons, electron-hole excitation or impurity states). These excitations could be accounted for by a perturbation theory of some kind, their dynamic nature imply that we would need to add one extra temporal dimension to $\hat{\tau}(r,r', t-t')$ which would eventually lead to so-called satellite peaks (subject of RIXS, but out of scope of this paper)\cite{Sebilleau-multi-ARPES-rev}. In an 'optically' dense 3D material usually only the nearest neighbour multi-site events would be accounted for, but recently significant progress has been made towards more accurate description from analytic\cite{Sebilleau-multi-ARPES-rev, Osterwalder-multiple-scattering} and numerical\cite{Dauth-TDDFT} approaches. 

Among various intrinsic and extrinsic looses the scattering path operator should also include vertex corrections (VC) that are due to a superposition of electronic waves escaping through various possible paths related through (nearly) elastic collisions. Following a seminal paper by Chang and Langreth\cite{Langreth-theor} (on ARPES theory extensions beyond sudden approximation\footnote{please note that in the formalism of Ref.\cite{Langreth-theor} , in TLL all the spectral function's weight is shifted to collective modes hence a probability of coupling between the dressed photo-hole propagator and the vertex plasmon is unitary}) these VC can be classified as a variant of interference terms and they are expected to be particularly relevant in the presence of degenerate states. While degenerate states are an exception in most 3D materials, the quasi-1D systems are special, since there are numerous degenerate states (each one defined in a different 1D chain) that may be equally good sources of a photo-electron. These are all elastic scattering events, hence they are singularly relevant perturbations and must be accounted for in the first place.  These wave interference processes are of main interest in section Sec.\ref{sec:Lommel}. For these VC the scattering path operator can be written as a geometric series:
\begin{equation}\label{eq:geom-final}
	\hat{\tau}(r,r')=(1-\hat{T}G_f(r,r'))^{-1}
\end{equation}
which contains information about emitted electron waves' propagation between TLLs -- here $G_f$ is an attenuated (by $\tilde{\Gamma}_{out}$) propagator of the wave between two 1D systems and $\hat{T}$ is the transmission/reflection operator that describes how the two waves superimpose). If the energy/momentum of the final states matches the initial state inside 1D wire then $\hat{T}$ is a number determined by wavefunction amplitude to the wire's boundary, if it does not match then final state propagates undisturbed by the presence of 1D wire (perfect transmission).  

%We will tackle these difficulties and construct the \emph{final state} in the specific quasi-1D case in Sec.\ref{sec:Lommel}.

\subsection{The emission process}\label{ssec:mod-proces}

According to Fermi's Golden Rule description of a photoemission intensity\cite{Feibelman-ARPESbasic-theo}:

\begin{equation}
I(E,k_{||})\sim \left[\sum_{\nu} \left\langle \Psi_f(E,k_{||})|\vec{A}\vec{p}+\vec{p}\vec{A}|\Psi_i(k,r)\right\rangle \delta(\sum_j E_j)\right]^2
\end{equation}
since we factorized the final state, it is tempting to factorize the initial state as well, that is to extract one given electron, as the one which undergoes the collision:
\begin{equation}\label{eq:ARPES-inten}
I(E,k_{||})\sim \left[\sum_{\nu} \left\langle \psi_f(E,k_{||})|\vec{A}\vec{p}+\vec{p}\vec{A}|\psi_\nu(k,r)\right\rangle \right]^2 \left\langle \Psi_{TLL}|c_k|\Psi_{TLL}\right\rangle \delta(\sum_j E_j)
\end{equation}
where $\delta(\sum_j E_j)$ imposes conservation of energy between final and initial states ($j=f,\nu$), the $\nu$ summation goes along all initial states that can contribute and $\psi_f$ is the final state. In stark contrast with common wisdom, based on a single-electron initial state, in the case of many body states there are many indistinguishable ways this extraction of the $\nu$-th electron can take place. The summation over $\nu$ covers not only internal degrees of freedom of a given 1D chain (these weights will sum up to produce spectral function of TLL), but it also goes along \emph{all} TLLs falling under the light spot (i.e. entire zone where $\vec{A}(r)\neq 0$). %If one now substitutes Eq.\ref{eq:final-def} with $\phi_l(r-r_i)$ taken as atomic (or molecular) orbitals, then (for properly chosen position of detector) the dipole matrix selection rules may be obtained. These do play a substantial role in ARPES intensity, but they are not the main focus of this work, as we are searching for other possible suppression factors.   

The separation of initial,final and intermediate states, although enables for an intuitive understanding of ARPES process, is very much arbitrary. One can realize that upon analysing the phases of the respective wavefunctions, that are profoundly connected with each other. Firstly, upon extracting one single-particle state from the total many body state, the phases of two constituents remain entangled. Secondly, a collision of the chosen single-electron with photon, is fully determined by a phase of this quantum mechanical scattering. Thirdly, even if the phase of photoelectron created in the collision is modified by the local part of $\Gamma$ (which is likely since there are many low energy electron-hole excitations) this may be still absorbed into a collision's phase by a local gauge transformation.   

%Clearly, see Eq.\ref{eq:ARPES-inten}, the key quantity is how the $\psi_f(\vec{r})$ overlaps with initial states of TLLs, hence in our per se inhomogeneous system it is crucial to find the space distribution of the final state. The final state is an electronic wave that is measured in the detector and its time evolution retraced back into bulk of the sample where a photo-emission took place. On its evolution the $\psi_f$ may undergo typical inelastic scattering/absorption on various states present in the solid (e.g. phonons, electron-hole excitation or impurity states). These excitations could be accounted for by a perturbation theory of some kind. However quasi-1D systems are special, since there are many degenerate states (each one defined in a different 1D chain) that may be equally good sources of a photo-electron. These are all elastic, hence they are singularly relevant and must be accounted for in the first place. 

\section{Disorder influence on spectral function}\label{sec:disord}

\subsection{Backward scattering}

An obvious issue, that substantially obscure our abilities when probing any fermionic propagators is disorder. The 1D systems are special in this respect and this can be illustrated by the following, qualitative reasoning. When one probes with ARPES a standard material with 3D dispersion then the light-absorbing electronic state is delocalized over entire spot of light on the sample's surface. For instance if one probes states in the vicinity of the $\Gamma$ ($\vec{k}=0$) point then a uniform response of entire spot is measured i.e. the initial state of photo-emitted electron is uniformly spread over entire spot. Since $k_{\perp}$ is a good quantum number, the electrons can move coherently around the obstacles i.e. impurities can be circumvented or (up to some extend) self-averaged. 

If we now insert an impurity inside a 1D electronic liquid, we immediately realize that electrons have no chance to circumvent it and propagate further. Instead they are back-scattered and their spectral weight is transferred into a localized state. Quantitatively, in the weak disorder limit, this can be expressed in terms of a renormalization group argument. While in higher dimensions impurities are either irrelevant or marginal, in 1D they are always a relevant perturbation and a collective disorder $D_b$ is a violently relevant perturbation with a large scaling dimension (the critical $K_{\rho}^c = 3/2 $ is larger than for any standard perturbation) \cite{giamarchi_book_1d}. Hence during RG flow the $D_b$ very quickly (especially for $K_{\rho}\ll 1$) reaches substantial values $D_b[l\approx \Lambda_0/O(1)]\rightarrow O(1)$ This implies that 1D chains that host disorder are strongly affected and removed to the localized sub-set already at pretty high energies. 

\begin{figure}[ht]
  \includegraphics[width=0.8\textwidth]{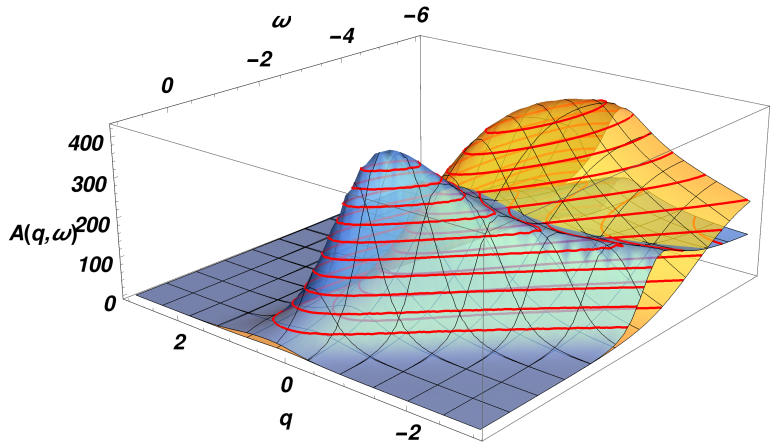}
	\caption{Comparison of single-particle spectral function $A(q,\omega)$  without and with (back-scattering) disorder as a function of energy $\omega$ (we took $V_F=1$ and $V_{\rho}=3$, zero energy at $E_F$) and momentum $q$ (measured in $\pi/a$ units). Blue surface (without disorder) has double maximum that reveals spin/charge separation, while yellow surface (with disorder) has only one broad feature. Disorder enters through renormalized $K_{\rho}$ parameter. %Righ: same like before but now for the amplitude of the holon peak
	We choose initial $K_{\rho}=1/3$ that is representative for a system with long range interactions and then follow its renormalization along the trajectory $K[l]=K_0 + D_b ArcTan[3/2 \pi/l]$ (the solution of the BKT flow) , where we took bare $D_b=0.05$. In experiment on disorder material one expects to observe a sum of the two surfaces.}
	\label{fig:Comparison}
\end{figure}

To illustrate it in Fig.\ref{fig:Comparison} we compare spectral weight of TLL(at $\omega$) and TLL with disorder where $\tilde{K}_{\rho}[l]=K_{\rho}-\frac{u_{\rho}}{u_{\sigma}}K_{\rho}^2D_b[l]$. The smaller the effective $\tilde{K}_{\rho}$ (i.e. the more distant from $\tilde{K}_{\rho}=1$, since we assume repulsive interactions $\tilde{K}_{\rho}<1$) the larger is the value of Green's function exponent $\alpha$. In practice this means that spectral weight available for ARPES close to $E_F$ is substantially diminished. Furthermore, in most cases shown above, the system turns from $\alpha<1/2$ to $\tilde{\alpha}>1/2$ i.e. the spinon peak turns into a threshold. For the holon peak singularity it is known\cite{Imambekov-10} that it is given by the power law $(\omega-v_{\rho}q)^{-\mu_+}$ where the exponent $\mu_+=-1/2 -(K_{\rho}+K_{\rho}^{-1})/4 + (K_{\rho}^{1/2}+K_{\rho}^{-1/2})/\sqrt{2} $, so singularity becomes weaker as $K_{\rho}$ decreases (and this is even without accounting for the fact that higher energy holon peak is always more prone to be broadened by decay into lower energy electron-hole states e.g. located on impurities). Overall spectral weight diminishes and shifts from peaks towards the incoherent part.  

\subsection{Forward scattering}\label{sec:forw-dis}

The situation in 1D systems is in fact even more complicated. It is not only the backward scattering component of disorder that excludes some 1D chains from the sub-set of 1D chains where TLL can be detected. The forward scattering disorder, usually assumed to be completely unobtrusive, in 1D case also has the ability to diminish the ARPES signal. 

The pertinence of forward scattering component of disorder is in fact a very fundamental difficulty. The forward scattering comes usually from more distant but charged inhomogeneities that act by means of long range Coulomb interactions. This is particularly relevant on the surface, where such interactions are only partially screened. With this respect a surface sensitivity of ARPES becomes a serious drawback.  

For the gapless TLL the forward scattering disorder can be solved exactly \cite{giamarchi_book_1d}. In bosonization language the term reads:
\begin{equation}
	H_f= \int dx \eta(x) \nabla\phi_{\rho}(x)
\end{equation}
(where $\eta(x)$ is a random distribution with a width $D_f$) and can be absorbed by a shift of $\phi_{\rho}$ field:
\begin{equation}
	\tilde{\phi}_{\rho} \rightarrow \phi_{\rho} + \int^x dx'\left(\eta(x') \frac{K_{\rho}^{-1}}{v_{\rho}}\right)
\end{equation}
Then any correlation function that involves $\phi_{\rho}$ field, including the single particle Green's function $G(x,t)$ (whose Fourier transform is measured in ARPES)is affected by this shift:     
\begin{eqnarray}\label{eq:forward}
	G_{R,L}(x,t)= \langle\exp\imath(\tilde{\phi}_{\rho}(x,t)-\tilde{\phi}_{\rho}(0,0))\rangle\langle\exp\imath(\phi_\sigma(x,t)-\phi_\sigma(0,0))\rangle\cdot \nonumber \\
	\langle\exp\pm\imath(\theta_\sigma(x,t)-\theta_\sigma(0,0))\rangle\langle\exp\pm\imath(\theta_\rho(x,t)-\theta_\rho(0,0))\rangle\\
	= \exp \left(- 2 \left[\frac{K_{\sigma-}^{-1}}{v_{\sigma-}}\right]^2 \int^x dx' \int^{x'} dx'' D_f \delta(x'-x'') \right) \cdot \nonumber \\
	\exp(\sum_\nu \langle\phi_\nu(x,t)\phi_\nu(0,0)\rangle\pm \langle\theta_\nu(x,t)\theta_\nu(0,0)\rangle)\\
	=\exp(-D_f |x|)A_{TLL}(x,t)
\end{eqnarray}
where $A_{TLL}(x,t)$ is a well known spectral function of TLL\cite{Orgad-TLL-spectr} expressed in terms of power-laws with exponents given by $(K_{\nu}+K_{\nu}^{-1})$. The second equality above comes from averaging over uncorrelated disorder. This spectral function is attenuated by an extra exponential factor.

There are two implications of this result: i) the presence of forward scattering disorder induce one additional source of intrinsic loses. This will reduce the penetration depth of photo-electrons escaping towards the detector; ii) upon taking Fourier transform of Eq.\ref{eq:forward} we see that the electrons (that succeed in reaching the detector) will acquire an additional Lorentzian broadening of their MDCs. Please note that our derivation is valid for real as well as complex $D_f$. The only restriction comes when the model is defined, as imaginary part of $\eta$ would lead to non-hermitian hamiltonian, something one avoids when a ground state is the aim. However in our stationary state, the $\psi_f$ component does follow an evolution given by a non-hermitian hamiltonian, hence the global constraint is removed. We will come back to this in Sec.\ref{sec:rand-Lomel}.

\section{The \emph{final state}: interference effects }\label{sec:Lommel}

Clearly, in Eq.\ref{eq:ARPES-inten}, the key quantity is how the $\psi_f(\vec{r})$ overlaps with initial states of TLLs, hence in our inhomogeneous system (1D chains and voids in between them) it is crucial to find the spatial distribution of the final state. We have effectively assumed (in Sec.\ref{ssec:mod-initial}) that each 1D wire can be modeled as a circular antenna emitting and reflecting electronic waves. Furthermore, since the \emph{initial} state electrons cannot propagate coherently between 1D systems, the emission event is strictly local on the perpendicular plane. However, when more than one indistinguishable trajectory is available, in quantum mechanics we are obliged to superpose all such possible trajectories. In the TLL with small $K_\rho$ (which is usually the case in systems with long range interactions) the single particle spectral weight is severely diminished at low frequencies which implies particularly low probability of inter-1D recombination of photo-carriers. %(here we also assume that only 1D states are available close to $E_F$). 
Furthermore, since electron-electron interactions strongly prefer forward scattering, this should also include the photo-electron (when it overlaps with TLL) or at least its momentum component along the b-axis (hence a component $\sim \cos \alpha$). This, in effect, brings $\cos \alpha$ vertex correction factor to propagation of the \emph{final state wave} which resembles the "optical vertex correction" in Fresnel diffraction [treatment of e.g. Hubbard model is still possible, but will be slightly more complicated]. 

%Hence we encounter a problem  of wave superposition phenomena in close analogy with such as e.g. diffraction on the sample.  
%To be precise, in this entire complex program, 
%\subsection{Case of a complex unit cell: Fresnel diffraction}
%This separation simplifies greatly a construction of the \emph{final} state $\psi_f$ and is justified by the peculiar properties of quasi-1D material. The fact  Also the relative orientations of atomic orbitals is a secondary factor as the final state waves care only about the relative distances between TLLs. 

All these assumptions point toward the Fresnel diffraction framework to describe electron waves inside the sample. We use a well known procedure where one inverts the time axis and solve a problem of the \emph{final} ARPES state by analyzing the LEED outcome, where a detector now plays a role of a (distant) source of an electronic wave. The wave falls onto the quasi-1D sample and now the 1D systems play a role of a cylindrical obstacle (the Babinet principle). We wish to know what is the resulting total wave-function inside the sample. The idea of using electronic wave diffraction to obtain information about ARPES wave-function is relatively well known [PES textbook], however very recently it has gained significant attention as it was experimentally shown that these effects may produce observable variations of intensities either with energy of incoming phonon\cite{Osterwalder-multiple-scattering} or even to reconstruct shapes of molecular orbitals involved in the process\cite{Puschnig-plane-wave, Weiß2015}.  

In the context of quasi-1D material, a further advantage of our simplified approach is that it admits an exact analytical solution. The source is distant so incoming waves can be taken as plane waves and paraxial approximation holds at least on the source side. The ARPES detector (LEED source) is always aligned along b-axis (we aim to measure this component of momentum since only this one is a good quantum number) so we know that along the b-axis the solution is simply a plane wave with periodicity set by $k_b$. In the perpendicular plane we consider a superposition of Fresnel diffraction patterns: we wish to know the amplitude of a wave elastically scattered on one 1D system in the position of another 1D system.   

The solution for the Fresnel diffraction on a circular aperture can be expressed in terms of Lommel functions\cite{Hufford-Lommel-orig}:
\begin{eqnarray}\label{eq:Lomm-sol}
\psi_{int}(r)=(\sin(N_{\bar{F}}^2(1+(r/R)^2)/2) + U_1(2 \bar{W}^{-1} N_{\bar{F}}, 2 \bar{W}^{-1} N_{\bar{F}} r/R))-\\
 i (\cos(N_{\bar{F}}^2(1+(r/R)^2)/2) - U_2(2 \bar{W}^{-1} N_{\bar{F}}, 2 \bar{W}^{-1} N_{\bar{F}} r/R))
\end{eqnarray}

where $R$ is the radius of the aperture (size of 1D conductor in our case), $r$ is a distance within perpendicular plane,  $U_{1,2}$ are Lommel functions of two arguments, $N_{\bar{F}}=R^2/(\lambda\bar{z})$ is the Fresnel number, $\lambda$ is a wavelength of a photo-electron inside the sample (directly proportional to square-root of photon energy) and $\bar{W}$ is a rescaling that accounts for a possible deviation from the paraxial approximation [cite]. The $\bar{W}>1$ when $\bar{z}$ is small, so the two factors tends to cancel each other. The Lommel functions have a damped (weakly aperiodic) oscillatory behavior.

We now proceed to construct the final state $\psi_f(r)$ along the following lines. Each 1D system (with radius $R$) scatters electronic waves and hence becomes a source of a wave $\psi_{int}(r)$. By superposition principle we add waves scattered by all 1D systems, this sum gives us $\psi_{f}(r>R)$. For the core part of the \emph{final state}, $\psi_{f}(r<R)$, we assume a confinement within a parabolic quantum well which gives a Hermite polynomial solution for a wavefunction. The amplitude of this part is fixed by a boundary condition at $r=R$. The procedure has to be applied self-consistently: we write Hermite polynomials with given amplitudes which determines strength of each antenna at $r=R$, then we propagate the partial waves (Eq.\ref{eq:Lomm-sol}) using superposition rule Eq.\ref{eq:geom-final}, which in turn gives us a new boundary condition (see App.\ref{ssec:DOC-bound} for details).   

\begin{figure}[ht]
	\centering
	\includegraphics[width=0.8\textwidth]{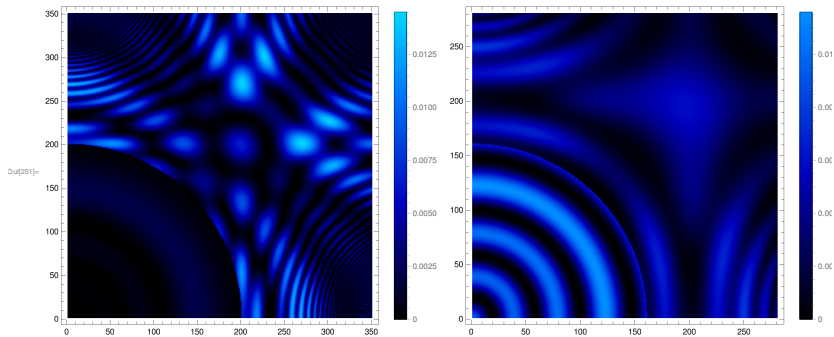} 
	\caption{Comparison of total \emph{final state} for two energies of photon energies. The choice on the right supports constructive superposition between nearest neighbours, the choice on the left is for de-constructive superposition at the 1D system boundary. To focus on detail only a quarter of a unit cell is shown, since the problem is defined on the square lattice reconstructing entire $\psi_f(r_\perp)$ is trivial (distance in a.u.).}
	\label{fig:FinalState}
\end{figure}

From this one can built an argument based on destructive superposition of electronic waves. %Consider \emph{i-th} TLL that shall give a strong electronic wave along its axis. [This is due to a coherent propagation of electrons along the b-axis, but not in a disagreement with diffraction construction where it corresponds to the Arago spot.] 
When each neighbour of a given 1D chain produces the electronic wave-function that is in anti-phase at the boundary, then the boundary condition value is reduced and an overall superposition of electronic waves inside ($ \equiv |\psi_f (r<R) \rangle $) is significantly reduced. The condition for this \emph{accidental de-selection rule} is that the distance between 1D systems is approximately two times larger than $\lambda$ (naturally there are also secondary \emph{de-selection} conditions corresponding to further minima of Lommel functions, but these are both weaker and harder to realize experimentally). This condition depends on the photo-electrons' momentum $\lambda^{-1}\sim k_0 $, hence the kinetic energy of photo-electron inside the sample $k_0 \sim \sqrt{E_{kin}}$, hence the effect can suppress the ARPES image for certain energies of incoming photons $E_{photon}=E_{kin}$. This is clearly visible in Fig.\ref{fig:FinalState} where for two different choices of wavelength of photo-electron we obtained drastically different distribution of \emph{final} state within the perpendicular plane. We can either enhance density in areas where TLLs reside or shift it towards void space in between them. 

\section{The \emph{final state} in the presence of disorder}\label{sec:rand-Lomel}

The problem with the disorder is even more profound than we explored so far in Sec.\ref{sec:disord}. Consider a charge neutral lattice distortion/dislocation that interacts only with phonons (and possibly some localized electron-hole transient states). According to our previous reasoning in Sec.\ref{sec:disord}, TLL should be immune at least to these (most common) perturbations. Unfortunately this is not the case. Any randomness in scattering rates, for instance due to coupling between electrons and phonons, when phonons propagate in a random environment, will then translate in the imaginary $\eta_{\Gamma}(x)$. Here the only assumption we make is that the $|\psi_f\rangle$ is always \emph{locally} in-phase with the \emph{initial} state present inside 1D system. To be precise: while e.g. electron-lattice scattering events are random, due to large velocity ratio of electrons and phonons these events average out to Brownian noise. What may happen is that for instance a lattice distortion will affect phonon dispersion and this in turn will cause an inhomogeneous (in space) Brownian noise. Again, electrons moving coherently in 3D would have averaged over these domains, but 1D electrons do not have this ability. This leads to inhomogeneous scattering $\tilde{\Gamma}(r_{\perp})$ and hence random $\eta_{\Gamma}(x,r_{\perp})$.   

We are now in the position to combine results of Sec.\ref{sec:forw-dis} and Sec.\ref{sec:Lommel} , and investigate the total $|\psi_f\rangle$ in the case when the forward disorder affects the 1D systems. Following previous reasoning random $\eta_{\Gamma}(x,r_{\perp})$ again leads to an exponential factor, this time it is a phase shift:
\begin{equation}
	G_{R,L}(x,t)= \exp(-D_{f\Gamma}(r_{\perp}) |x|)\exp(\sum_\nu \phi_\nu(x,t)\pm\theta_\nu(x,t))
\end{equation}
Note that this depends on $ r_{\perp}$, so we admit that the phase shift may be different for different 1D systems. Let us come back to the argument put forward in the introduction, that is a photo-emission from quasi-1D material taken as a superposition of electron-waves emitted from separate antennas. What we have established just now is that in the presence of forward scattering each of these antennas acquire a different phase. From this we deduce that the phases of electronic waves emitted from various 1D systems are now randomly distributed. This is in close analogy with optical waves interfering through apertures with randomly added phase-shifting plates, an optics problem with a textbook solution -- we expect that superposing such signals will produce a strongly suppressed net outcome. 

\begin{figure}[ht]
	\centering
	\includegraphics[width=0.8\textwidth]{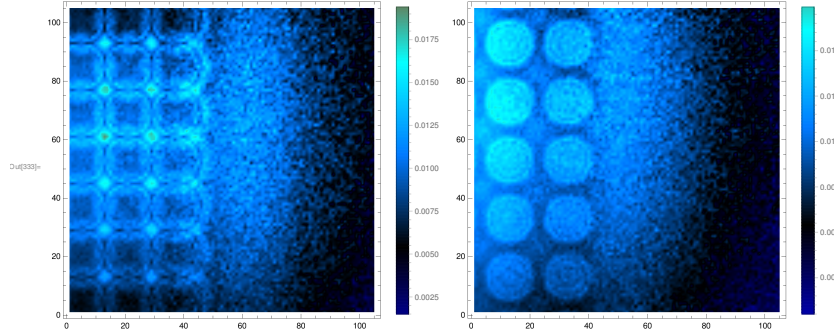}
	\caption{Comparison of total \emph{final state} for two energies of photon energies, but this time with disorder. Again the choice on the right supports constructive superposition between nearest neighbours, the choice on the left is for de-constructive superposition.}
	\label{fig:FinalState-disord}
\end{figure}

In Fig.\ref{fig:FinalState-disord} we show the damaging effect of phase-disorder. On the left of each picture we have an unperturbed system with a clearly visible square lattice of 1D systems. As we moved towards right the random phases in between 1D systems are introduced. We can observe a drastic drop of the \emph{final state} amplitude on these chains on the right on each panel.

%[Note: There is also plot with disorder proportional to $\cos$ where periodic suppression is visible, but then I do not have such a nice picture of undisturbed crystal. Probably perturbed zones are to close to each other]

%\section{Case study: NbSe3}

%\subsection{Col.C}

%\subsection{Col.A}

\section{Discussion: implications for experiments}

One key question is whether the results obtained above, in particular in Sec.\ref{sec:Lommel}, can lead to experimentally significant effects. Firstly we note that when the photoelectron's wavelength in not too small and inter-1D wires' distance is not too large (they are the same order as $R_0$) then the function $\psi_{int}(r)$ in Eq.\ref{eq:Lomm-sol} varies slowly which implies that solution is stable with respect to small variation of geometry (including variations of $N_{\bar{F}}$). This conjecture is possible thanks to a direct link with established results from the field of optics. Furthermore we employ this link, to find that Fresnel diffraction pattern stays stable both for soft-edges 'obstacle'\cite{rough-Fresnel} as well as elliptical apertures\cite{Borghi-eliptFresnel} (in the latter case one even observes the amplitude oscillations with stronger amplitude). Hence the postulated effects should be valid in real material. 

Peculiar conditions of photo-electron scattering in a quasi-1D material makes it possible to derive conditions of constructive/destructive interference %The disorder will enhance the destructive interference, which proves its pertinence on this even more profound level. In the pure sample case
, these conditions depend on energy of incoming photons as well as experiment geometry. From the practical view this means that for certain energies of photons the band dispersion will be quite discontinuous, in practice making it impossible to detect unambiguously. Hence a negative result in detecting the 1D band does not necessarily imply that 1D sates are not present, as it may be rather a result of the "accidental de-selection rule" emerging from the interference vertex corrections. One way to experimentally disentangle this phenomenon (apart from changing energy of photons, which is sometimes impossible) is to check how strongly band's intensities vary in different Brillouin zones (a destructive superposition will be slightly different when $k_b$ modulation changes). Here, we wish to point out that the "diffraction" effect proposed here may also cause ARPES intensity variation as a function of $k_\perp$. It is rather counter-intuitive effect, where variation in the perpendicular direction is in fact an evidence of 1D states.

The simplicity of our approximation allows for a rather quick verification if indeed a conditions for the "accidental de-selection rule" were met. On the positive side, this paves way to design experiments in a way to harvest constructive superposition and enhance our chances to probe TLL. Certain conditions must be met to validate our reasoning:
\begin{itemize}
\item 1D system must be in a gapless state and away from a dimensional cross-over. In particular, the gapped TLL would allowed for low-energy in-gap states (solitons, breathers) which could scatter/absorb photo-electrons
\item only 1D carriers are available close to $E_F$. This is again to avoid further scattering/absorption events, please note that it may be sufficient to separate other states by means of symmetry selection rules
\item the wave-length of photo-electron should be similar to characteristic dimensions of the material (which usually means UV-ARPES of ultra-soft X-ray ARPES). This is also regime where the sudden emission approximation is most likely violated  
\end{itemize}   
The last point raises the issue how in practice to determine e.g. the effective radius $R$ or the inter-chain distance. Obviously, this depends on the material under consideration, with a general rule that if there is a coherent hopping between two sites then they should be considered as a constituents of a single unit ("aperture"). On the other hand, two degenerate bands, suggesting a presence of a doublet (of e.g. columns) with no hopping implies that these two constituents should be taken separately. Finally, we mention that to achieve such an atomistic addressing of photons one should be able to use light polarization and/or vary photon energies (various atoms will have orders of magnitude different cross-sections). These can make our study viable for experimental test. 

\begin{figure}[ht]
  \centering 
  \includegraphics[width=0.45\textwidth]{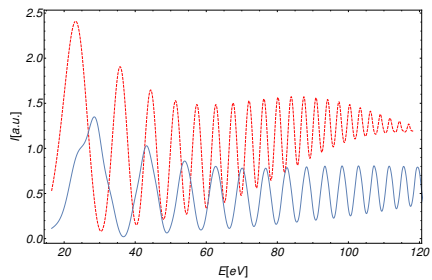}
	\caption{Comparison of spectral intensities coming from initial states on 1D wires of different sizes, shown as a function of incoming photon energy. We took dimensions corresponding to situation in $NbSe_3$: for a so-called col.A (red) $R=3.3A$ and $d=1.9A$ , and for col.C (blue) $R=0.75A$ and $d=3A$.}
	\label{fig:Col-Compar}
\end{figure}

One particularly interesting material, for experimental confirmation of our theory is NbSe$_3$ (and its analogues, columnar tri-chalocogenides) where different bands correspond to columnar structures of different sizes. NbSe$_3$ consist out of several columns with different relative orientation in space, but identical chemical content. The arrangement is such that some (pairs) of the columns are strongly hybridized, which gives rise to structural units with $R_0 \approx 3A$, while other columns are not hybridized, which produces structural units with $R_0 \approx 1A$. The bands originating from electrons residing on different types of columns can be easily differentiated, as their dispersions differ by $\approx 0.2eV$. This offers an ideal setting to test if relative intensities of the two types of bands will vary with energy of incoming photons as illustrated in Fig.\ref{fig:Col-Compar}. %Indeed, by comparing [] and [], we note that such effect has been likely observed.

\section{Conclusions}

There are two main findings of this paper. Firstly, when one attempts to probe the TLL by ARPES spectroscopy, it is of an uttermost importance to avoid any sources of disorder as any form of disorder can harm the outcome much more substantially than in standard 3D materials. Secondly, we focused on a stage of ARPES process that is usually overlooked -- the propagation of a photo-electron towards the surface. It has been shown that in the case of quasi-1D material this process may be quite complex and supports quite rich structure. Electronic waves' superposition effects can cause an order of magnitude changes in intensity. Finally, we put these two aspects together and show that forward disorder leads to a destructive superposition. Among other implications, the one most important is that in a quasi-1D material disorder is able to act non-locally -- when a disordered 1D system is within a photon spot not only it will not contribute, but it will also severely distract signal from neighboring chains. It is then highly desirable to probe a smaller spot on the sample if we can ensure that this is a region of the highest purity. 

Finally, the situation of 'destructive' interference which result in larger amplitude of $\psi_{f}(\vec{r})$ in between 1D wires, in real material (where there are actually some other atoms in between 1D systems) may result in an enhanced probability of the MERS processes -- a kind of multi-site emission that is recently under an intense scrutiny. The results of our work, Sec.\ref{sec:Lommel}, should be useful also in this context.  

\appendix

\section{Relation to other formalisms} 

%One peculiar case, where it is hard to justify the above given homogeneous circular aperture construction is when the 1D chain consist on one sole orbital (e.g. protected by symmetry and large energetic distance from other atomic orbitals). For pure single orbital case with one atom in a unit cell, the electronic wave-function can have only one orbital character for each momentum, there is no possibility of mixing effects due to electron-electron interactions\footnote{please note: for the case of nanotubes there is one sole orbital, but the eigenstate is distributed along many sites (around tubes circumference) and each of these has its own system of coordinates (curvature effects), so the "circular aperture" construction does hold in this case}. In such a case, instead of considering the Lommel functions one should express the wave emitted from each TLL as a Bessel function $\psi_f(r) \sim J_l(r/R)$ (times an $l-th$ spherical harmonic given by the angular momentum of the atomic orbital), such form will hold between the nearest neighbours (at larger distances, the evanescent nature of the wavefunction, encoded in the Hankel function component will gain more weight). Needless to say, the Bessel function also has the oscillatory behaviour, so the previous argument holds, that has lead us to the \emph{accidental de-selection rule} still holds.

A high energy \emph{final state} electron is a nearly free particle inside the material (or at least in the space in between 1D chains) and naturally for this case the solution of Schrodinger equation is known. To be precise two families of solutions were obtained. The best known is in terms of plane waves, this allows for straightforward boundary conditions on the crystal surface. The presence of this function can be immediately separated out in Eq.\ref{eq:Lomm-sol}. The other solution allows to match easily with centrosymmetric potential(s) and in 2D is expressed in terms of cylindrical Bessel functions of first and second kind. The Lommel functions themselves can be expressed as a sum of Bessel functions:

\begin{equation}\label{eq:Lom-to-Bes}
U_n (w,z) \approx \sum_{m=0}^{\infty} \left(\frac{w}{z}\right)^{n+2m}J_{n+2m}(z)
\end{equation}

Clearly, the solution used in this paper is a particular case that matches best the boundary conditions we have imposed. Ultimately, the minimal ingredient of the model that is required is the presence of a different (complex number) potential inside 1D chains ($r<R$) and in between them ($r>R$). By taking a limiting behaviour of the Bessel function of the first kind $J_{\nu}(z)\approx (z/2)^{\nu}/\Gamma(\nu+1)$ (when $z<1$, here $\Gamma(\nu)$ is a Gamma function equal to factorial for integer $\nu$ showing that the sum in Eq.\ref{eq:Lom-to-Bes} converges quickly) and $J_{\nu}(z)\approx \cos(z - \nu\pi/2 - \pi/4)\sqrt{2/(z\pi)}$ (when $z>\nu^2 - 1/4 $) we see that Lommel functions $U_{n}$ saturates to a constant value for small distances, while away it gives a cylindrical (decaying) waves with various phase shifts, that may be interpreted as signals from various scatterers inside the ''aperture''. If one had a knowledge about distribution of scattering phase shifts for TLL in a given material, then he/she should modify the definition of Lommel function accordingly. This establishes a link between our "circular aperture" approximation and eventually the full atomistic solution. %Now, noticing that in our previous "Fresnel-diffraction" argument we have effectively attempted to capture a sum of several atomic orbitals the single orbital limit is naturally a case with only one term in the sum Eq.\ref{eq:Lom-to-Bes}.

Concerning possible generalizations of our formalism, one has to keep in mind that in here we have considered only scatterers (1D chains) in s-wave channel. An extension to higher (angular) harmonics would require several modifications of the formulas. For the state inside the 1D chain, instead of Hermite polynomial we would need to use Laguerre polynomial -- a full solution of 2D harmonic potential (Hermite polynomial is a special case of it). For the (nearly) free electron solution in between the chains, Eq.\ref{eq:Lomm-sol}, we would need to multiply it by a linear combination of spherical harmonics (we assume that the solution can be factorized) and the sine/cosine terms would need to be supplemented by further spherical Bessel functions (they represent radial part of scattered waves). The extension to higher harmonics should be done with extra caution though: a great advantage of constraining our calculation to the s-wave in-chain state is that then a boundary condition is effectively given by an average at $r=R$ (see App.\ref{ssec:DOC-bound}) which does not depend much on details of calculations (while the precise distribution of amplitude's variations along the boundary, at $r=R$, may change substantially).

\section{Details of the boundary condition for $\psi_f$}\label{ssec:DOC-bound}

As usual in the first quantization problem the key issue is to find a wavefunction that will fulfill the continuity condition on the boundary. A difficulty in our case is that the wavefunction outside a given 1D chain is a superposition of Lommel waves scattered on all other chains, hence it will in general not obey the s-wave symmetry. We then minimize an integral $\int Abs(\psi_{in}(R_w)-\psi_{out}(R_w)) d\alpha$ where angle $\alpha$ is a cylindrical coordinate and we take an integral over the cylindrical boundary at a given $R_w \approx R_0$. A parameters that we can vary are amplitudes of each wavefunction and the radius $R_w$ (a soft-boundary model). We then optimize for s-wave component, which is justified by the assumption about the s-wave character of the \emph{initial state} (and by the fact that the Laguerre polynomials forms orthonormal set, which constrains \emph{final state} through Eq.\ref{eq:ARPES-inten}).

One could argue that use of different basis set could alter the above result and% Please also note that I restricted myself only to s-wave \emph{final} state, which may be justified for emission from s-wave or $d_z^2$ orbital initial state, including other in the trial state is also a possible extension. One could also discuss 
the precise way the boundary condition is defined (we choose a stationary point).%, minimum, of amplitude and derivative mismatch). 
However all these, while important quantitatively, does not affect the main result: if there is a negative superposition on the 1D system boundary at $r=R_W$, then amplitude of the \emph{final} state inside (for $r<R_w$) is reduced.
A careful inspection of Fig.\ref{fig:FinalState} reveals that there is always a maximum of Lommel function for $r$ a bit above $R$ which naively ca lead to assertion  that if one takes slightly bigger $R_w$ then this will increase the amplitude inside 1D system. However this is not true as this maximum is actually movable i.e. when $R_w$ increases the maximum moves away as well. Furthermore, one notices a fine structure of fringes visible in our solution $\psi_f(r)$, but these details may be unphysical. It is enough to account for electron-electron interactions (on the simplest level, sufficient for the (sparse) photo-electrons), for instance the exchange interaction on GGA level to be precise, to wash out all these fine details. However the broad distribution of $\psi_f(r)$ should stay intact.   

\section*{References}
\bibliography{biblio-Fresnel}

\end{document}